**Gaussian-splatting ptychography via explicit and interpretable primitives**

Qianhao Zhao[1,†], Zhixuan Hong[1,†,*], David Brady[2], Changhuei Yang[3], Andrew Maiden[4,5], Zhongtian Zheng[1], Ruihai Wang[1], Daniel Gage[6], Mary Lipton[7], Christopher Anderton[7], Arunima Bhattacharjee[7], and Guoan Zheng[1,*]

[1]Department of Biomedical Engineering, University of Connecticut, Storrs, CT 06269, USA
[2]Wyant College of Optical Sciences, University of Arizona, Tucson, AZ 85721, USA
[3]Department of Electrical Engineering, California Institute of Technology, Pasadena, CA 91125, USA
[4]Department of Electronic and Electrical Engineering, University of Sheffield, Sheffield, S1 3JD, UK
[5]Diamond Light Source, Harwell, Oxfordshire OX11 0DE, UK
[6]Department of Molecular and Cell Biology, University of Connecticut, Storrs, CT 06269, USA
[7]Pacific Northwest National Laboratory, Department of Energy, Richland, WA 99354, USA
[†]These authors contributed equally to this work
*Email: zhixuan.hong@uconn.edu or guoan.zheng@uconn.edu

**Abstract:** Ptychography overcomes the limits of lenses by co-designing optics and computation. Yet prevailing implementations reconstruct on a pixel grid, where weakly-constrained modes drift and recovery demands redundant data. Here we introduce Gaussian-splatting ptychography, representing object and probe as Gaussian primitives. Relocation concentrates primitives where structure is dense, and overlapping primitives couple neighbouring pixels to suppress mode drift. The scheme unexpectedly restores the low-frequency phase that conventional approaches lose, enabling uniform phase transfer across spatial frequencies. The probe is represented and updated in its pupil plane from a random start. In Fourier ptychography, the pupil-plane model recovers severe aberrations where pixel-grid solvers fail. In conventional optical, X-ray and electron ptychography, the pupil-plane primitives also recover the real-space probes with no model of beam-forming optics. The representation cuts memory up to 14-fold and recovers specimens from fewer acquisitions. At electron wavelengths, it resolves atomic structure at tens of electrons per square ångström.

## Introduction

Ptychography has become an enabling modality of computational imaging. It recovers the complex transmittance of a specimen from a series of intensity measurements acquired under laterally shifted illumination[1, 2]. The concept dates to 1969[3] and became practical with the iterative algorithms introduced in 2004[4]. By replacing the limits of physical lenses with a tractable inverse problem, ptychography has reached deep-sub-ångström resolution with electron[5, 6] and imaged integrated circuits non-destructively in three dimensions with hard X-rays[7, 8, 9]. At optical wavelengths it has grown from a microscopy technique into a general imaging platform[10]: it delivers gigapixel whole-slide imaging with both high resolution and a large field of view[11, 12, 13, 14], recovers label-free molecular contrast in the deep ultraviolet[15], sees through the bare tip of an ultrathin fibre bundle for endoscopy[16, 17], captures gigapixel videos of mesoscale dynamic specimens[18], and synthesizes apertures at long working distances[19, 20, 21, 22, 23]. Together, the operation of ptychography spans more than 9 orders of magnitude in length scale[24], from picometre electron wavelengths to macroscopic objects imaged at over a hundred metres.

Despite this diversity of wavelengths and geometries, prevailing implementations share a representational choice. The specimen is reconstructed as a dense complex-valued array, with one independent unknown assigned to every pixel of the reconstruction grid. The pixel is inherited from the detector; it is not chosen for the specimen. The number of free parameters is therefore fixed by the space-bandwidth product of the measurement system[25] and is blind to the content being imaged. An empty field of view costs exactly as much to represent as a densely structured one, and a gigapixel complex field occupies 8 GB in single precision before any optimizer state is counted, so recovery at this scale proceeds tile by tile[26]. Two further consequences follow. First, because every unknown must be constrained by the data, the parameterization creates a structural demand for redundancy. Adjacent measurements are typically overlapped by 50% or more to make the problem well conditioned[27]. This redundancy multiplies acquisition time, radiation dose and data volume, and it is particularly costly for beam-sensitive specimens[28]. Second, a pixel grid imposes no coupling between neighbouring unknowns. Every pixel is free to move independently, so weakly-constraint modes are subjected to drift. The probe is the clearest case. In blind ptychography, the probe is recovered jointly with the object, and each measurement constrains only their product, so the probe is weakly constrained until the object has taken shape. The same weakness leaves reconstructions sensitive to uncertainties in scan position, defocus and pupil aberration, and prone to stagnation in local minima.

These failure modes specify what a better representation should provide. Its parameters should follow the structure of the specimen rather than the sampling of the detector. Neighbouring points should be coupled, so that a weakly constrained mode is spanned by few parameters and cannot drift on its own. The parameters should be explicit and physical, so that they can be inspected and constrained from prior knowledge. And the representation should be differentiable through any forward model, so that object, probe and experimental parameters can be recovered together. Existing remedies meet some of these requirements and not others. Explicit regularization adds coupling to the grid[29, 30, 31], but the parameter count is unchanged, the prior is fixed before the specimen is seen, and a penalty weight must be tuned by hand for each dataset. Implicit neural representations decouple the parameter count from the grid[18, 32, 33, 34, 35, 36, 37, 38, 39, 40], but their weights carry no physical meaning and their resolvable bandwidth is set by architectural choices such as positional encoding rather than by the specimen; no parameter can be read as a position, a size or an orientation, so the representation cannot be inspected, edited or initialized element by element.

Explicit primitives meet the demands set out above. An anisotropic Gaussian is the minimum-uncertainty element jointly localized in space and spatial frequency. Every point of the field is written by several overlapping elements, so neighbouring points are coupled by construction. Every parameter is a position, a shape or a weight. And the composition is a sum, so it is differentiable through any forward model. Gaussian splatting in computer vision provides the machinery to optimize a large population of such elements from a random start[41, 42, 43, 44], and the same primitives have been fitted to holograms for display[45] and reconstruction[46]. Whether such elements would converge in ptychography was an open question. The difficulty lies in what ptychography measures. Every measurement is formed from the product of the object and the probe, and the detector records only the intensity of that product after propagation, so neither factor is observed on its own. Such partial views must agree on one object, and the phase to be recovered carries dense $2\pi$ wraps. The primitives therefore have to carry two coupled fields at once, with nothing but the data to tell them apart.

Here we introduce Gaussian-splatting ptychography. Both the specimen and the probe are represented as explicit anisotropic Gaussian primitives, each carrying a learnable position, covariance and complex weight. The weight is stored as real and imaginary parts rather than as amplitude and phase, so that the optimizer never meets the branch cuts that phase wrapping introduces. The primitives are splatted onto a regular grid, embedded within modality-specific forward models and optimized end to end by automatic differentiation. Relocation recycles underused primitives toward regions that carry the most signal, so sparsity becomes intrinsic rather than imposed. Unexpectedly, we find that two characteristic failure modes of grid-based recovery disappear. First, the low-spatial-frequency phase that conventional approaches lose is restored, so the reconstruction recovers absolute optical thickness rather than edge contrast. Second, blind recovery of severe, spatially varying pupil aberrations in wide-field optical microscopy succeeds where pixel-grid solvers fail, and because the probe of conventional ptychography is represented in its pupil plane, the same robustness carries over to X-ray and electron ptychography, where object and probe are recovered from a random start. We validate the framework on coded, Fourier and conventional ptychography, spanning optical, X-ray and electron wavelengths. At optical wavelengths, it delivers gigapixel reconstructions at one order of magnitude lower memory cost. At electron wavelengths, it resolves atomic structure from an ultralow-dose dataset at tens of electrons per square ångström, starting from a random pupil with no model of the probe-forming optics.

# Results

## Principle of Gaussian-splatting ptychography

In Gaussian-splatting ptychography, the complex transmittance of the specimen is not stored on a pixel grid. Instead, it is synthesized from a collection of explicit primitives (Fig. 1a). Each primitive carries a two-dimensional centre coordinate, an anisotropic covariance parameterized by two scales and an in-plane rotation, and a complex weight expressed by its real and imaginary components (Supplementary Note 1). The two components are splatted onto a regular grid by a tile-based rasterizer that returns analytic gradients for every primitive parameter, and are coherently summed to render the complex field (Supplementary Note 2). The rendered object is therefore a superposition of smooth, oriented footprints rather than a set of independent pixel values. The probe or pupil is carried by an independent primitive population with its own scale bounds and learning rate, so that specimen and probe are recovered together whether the probe is a smooth, low-order pupil aberration or a highly structured illumination beam.

An amplitude-phase parameterization of the weight would introduce branch cuts at 2π wrapping boundaries, which are non-differentiable and trap the optimizer[18]. The real-imaginary form describes the identical set of fields under an identical loss but removes the discontinuities from the parameter map (Supplementary Note 1). Supplementary Fig. S1 compares the two forms on a crystal, a bacterial microcolony and a thyroid smear. The real-imaginary representation preserves fine structure in the first two specimens and converges on the thyroid smear, for which the amplitude-phase representation fails entirely.

The rendered object and probe are propagated through a modality-specific forward model to predict the measured intensities, and the loss is evaluated against the raw measurement stack (Fig. 1b). We implement three forward models (Fig. 1c). In conventional ptychography, a localized probe is scanned across the specimen and the far-field diffraction pattern is recorded[4]. In lensless coded imaging, an extended beam illuminates the specimen and a translating coded surface close to the detector modulates the transmitted wavefield[12, 47]. In lens-based Fourier ptychography, angle-varied plane waves illuminate the specimen and band-limited images are collected through an objective, which synthesizes a numerical aperture beyond that of the objective[11, 48]. Every operation from primitive parameters to predicted intensity is differentiable: the rasterizer supplies analytic gradients for the primitives (Supplementary Note 2) and the forward model is differentiated automatically. Gradients therefore propagate back to the primitives and, at the same time, to experimental parameters. Supplementary Fig. S2 demonstrates joint scan-position correction in simulation. Position recovery improves the amplitude and phase peak signal-to-noise ratio (PSNR) from 23.65 to 35.47 dB and from 30.48 to 42.07 dB, respectively, and reduces the coordinate root-mean-square error from 1.62 to 0.16 pixels.

Optimization proceeds from a random initialization (Fig. 1d). At intervals, the primitives carrying the least weight are declared dead and respawned beside surviving primitives sampled in proportion to their weight, with the weight shared so that the rendered field is unchanged at the moment of relocation; the optimizer state of each relocated primitive is reset (Supplementary Note 3). This differs from the clone-split-prune heuristics of the original splatting formulation and follows instead the sampling-based reformulation of splatting optimization. A primitive that drifts away from useful support receives a vanishing gradient and stops contributing. That region is then under-represented, with no mechanism for recovery. Supplementary Fig. S3 shows precisely this outcome in coded ptychography, where reconstruction without relocation exhibits a localized convergence failure that relocation eliminates. In X-ray ptychography, relocation also improves the half-period Fourier-ring-correlation resolution at the 1/2-bit threshold from 22.7 to 14.1 nm (Supplementary Fig. S4).

Because the representation is explicit, its cost is transparent. Each primitive is stored as seven single-precision parameters: two coordinates, two scales, one rotation angle and two weight components. The storage cost of a reconstruction is therefore the primitive count multiplied by 224 bits, independent of the rendering grid. Expressed as bits per pixel (bpp) of the rendered field, this makes reconstruction fidelity and representation size directly comparable. Supplementary Figs. S5 and S6 sweep the primitive count from 1,000 to 100,000 on a 1,024 × 1,024 field for Fourier and coded ptychography, respectively, spanning 0.21 to 21.36 bpp. In both modalities the PSNR rises steeply and then saturates, with diminishing returns beyond roughly 8 bpp. For reference, an uncompressed dense complex-valued grid in single precision costs 64 bits per pixel, and we use this value as the baseline throughout. The gigapixel reconstructions reported below fall between 4.67 and 10.68 bpp, a 6- to 14-fold reduction in storage.

**Overlapping primitives restore low-frequency phase through uniform phase transfer**

In conventional Fourier ptychography, the phase transfer function is strongly non-uniform across the synthetic aperture. It approaches zero at the origin. Low-spatial-frequency phase is therefore only weakly encoded in the intensity measurements and is poorly recovered[49, 50]. Extended features are reconstructed with sharp edges, but their interiors sag toward the background, and absolute phase steps are systematically underestimated. A pixel-grid solver offers no remedy, because it leaves each pixel free to drift wherever the data are uninformative.

Figure 2 compares pixel-grid automatic differentiation with Gaussian splatting on four specimens that we imaged using our Fourier ptychographic microscope platform (Methods). Both methods use the same raw data, forward model, loss and optimizer. On a quantitative phase target of known step height (Fig. 2a), the automatic-differentiation reconstruction resolves the bar edges but recovers a step height well below the true value. The interior of each extended feature relaxes toward the background. Gaussian splatting reproduces the full step height and maintains a flat response across the feature, matching the ground-truth profile plotted in the inset. The same behaviour holds for cystine crystals and a mouse-kidney section (Fig. 2b,c). The line profiles show the automatic-differentiation phase collapsing between edges, while the Gaussian-splatting phase remains level across each structure.

The effect is most pronounced for optically thick specimens. In a $Na_2CO_3$ crystal preparation (Fig. 2d), the loss of low-frequency phase leaves the automatic-differentiation reconstruction with little more than the striation edges. Gaussian splatting recovers the accumulated optical thickness, which appears as dense $2\pi$ phase-wrapping fringes across the crystal body. The fringes are not an artefact of the representation. The crystals imaged by lensless coded ptychography (Supplementary Fig. S7) show the same wrapped-phase structure. In that modality the coded surface converts phase into intensity contrast without the low-frequency deficit of Fourier ptychography[12, 47]. Absolute optical thickness, not edge contrast, is what quantitative phase imaging is meant to deliver[49, 51].

To identify the origin of this behaviour, we measured the end-to-end phase response of each reconstruction pipeline directly, using a simulated binary phase object with a $\pi$ step (Supplementary Fig. S8). Conventional Fourier ptychography exhibits the expected V-shaped response that approaches zero at the origin. The Gaussian-splatting reconstruction maintains an approximately uniform response across the whole synthetic aperture. The measurements are identical in the two cases, so the representation adds no information; it determines how the weak low-frequency information in the intensities is used. On a pixel grid that information is overwhelmed by the freedom of each pixel to drift, whereas in the splatting representation every pixel of the rendered field is written by many overlapping primitive footprints, so a slowly varying phase offset is spanned by far fewer degrees of freedom and cannot decay toward the background while leaving the edges intact. The effect is not attributable to the loss function: the pixel-grid baseline was run with the identical gradient-domain loss, forward model and optimizer (Supplementary Note 4), so the two reconstructions differ only in how the field is represented.

**Robust blind recovery in gigapixel Fourier ptychography**

We next acquired a gigapixel Fourier ptychographic dataset of an H&E-stained tissue section on the same platform (Fig. 3a). The full-field colour reconstruction is stored by the Gaussian-primitive representation at 8.5 bits per pixel per colour channel, about one eighth of the 64 bits per pixel required by a dense complex-valued grid. Starting from a random initialization, the primitive map condenses onto the tissue architecture, and the rendered complex object evolves from noise to a high-contrast amplitude and phase reconstruction within ~20 epochs (Fig. 3b,c). The intermediate primitive maps are themselves interpretable: because the primitives carry explicit geometry, their density marks where the tissue is structurally complex and their covariances encode local orientation, so the maps trace glandular boundaries and stromal fibre orientation without any segmentation step. We do not pursue such analyses here, but no equivalent quantity can be read from a pixel grid without a separate pass over the rendered image.

The more demanding test is blind recovery of the pupil aberration. In a wide-field system the aberration varies across the field of view[52, 53], and it must be estimated jointly with the object from the same intensity data. Here each region is reconstructed on its own from a random pupil initialization, with no information transferred from neighbouring segments, no prior model of the field dependence, and no parametric aberration basis. Figure 3d,e compares three reconstruction schemes on an off-axis region, where aberrations are most severe. Quasi-Newton optimization[54] fails to converge to a physically meaningful pupil in any colour channel and returns a reconstruction close to the low-resolution raw measurement. Pixel-grid automatic differentiation recovers low-order aberrations in the green and blue channels but leaves residual structure in the red channel, and its colour composite shows pronounced artefacts. Gaussian splatting recovers the smoothly varying aberrations in all three channels and resolves glandular and stromal morphology with the best fidelity.

In Supplementary Fig. S9 and Supplementary Table 1, we quantified this behaviour in simulation against a known pupil. We compared four schemes with different aperture overlap: the extended ptychographic iterative engine (ePIE)[55, 56], quasi-Newton optimization, pixel-grid automatic differentiation and Gaussian splatting. At the highest overlap all four recover the object. The three optimization-based schemes perform comparably, with Gaussian splatting highest in both channels (39.80 dB amplitude, 47.43 dB phase). The pixel-grid schemes then fall away at different rates. Quasi-Newton degrades sharply below 78% overlap, whereas automatic differentiation holds to 58% before collapsing at 39%. At 39% overlap, ePIE, quasi-Newton and automatic differentiation all return pupil estimates dominated by spurious phase discontinuities. Gaussian splatting returns a smooth, low-order aberration consistent with its high-overlap result, and it exceeds automatic differentiation by 8.7 dB in amplitude and 9.2 dB in phase. The advantage therefore widens precisely where the data constrain the pupil least. Supplementary Fig. S10 repeats this four-way comparison on the experimental H&E histology dataset of Fig. 3, channel by channel; on these measured data the ranking is the same as in the simulation of Supplementary Table 1.

The aberration result suggests that the representation should also tolerate sparser sampling. We tested this on a blood smear imaged by Fourier ptychography at three sampling densities: 169, 49 and 25 raw images, corresponding to approximately 70%, 41% and 17% aperture overlap (Fig. 4). The lowest of these is far below the 50-70% overlap conventionally required for stable convergence[27]. At the highest density, both automatic differentiation and Gaussian splatting recover cellular morphology and a smooth pupil aberration. As the overlap decreases, the automatic-differentiation reconstruction degrades progressively. It fails to converge at 25 raw images, and its pupil estimate collapses into noise. Gaussian splatting preserves cellular morphology throughout. The pupil aberration it returns at 25 images remains consistent with the one recovered at 169. This self-consistency check is available even though the true aberration is unknown.

Supplementary Fig. S11 tracks both primitive populations for on-axis and off-axis regions of the smear over 40 epochs: the object primitives organize around cellular structures, and the pupil primitives converge to smooth aberration maps that differ between the two field positions, as expected for a spatially varying aberration, yet yield consistent cellular morphology at both.

**Gigapixel lensless coded ptychography with reduced acquisitions**

We next turned to the lensless coded modality, in which a translating coded sensor encodes sub-pixel information into the recorded intensity[12, 47]. We acquired 600 raw images of live bone-marrow mesenchymal stem cells on our coded-sensor platform, with the platform inside the incubator (Methods). Figure 5 shows the full-field phase reconstruction across a gigapixel field of view. The reconstruction is stored at 4.67 bits per pixel, roughly a 14-fold reduction relative to a dense complex-valued grid. As optimization proceeds, the primitives organize along cell bodies and protrusions, and the rendered phase sharpens correspondingly (Fig. 5a,b).

The reduction in required acquisitions is substantial. Using the 600-measurement reconstruction as the ground-truth reference, Gaussian splatting reaches a phase PSNR above 26 dB from a single raw image and rises to approximately 35 dB. Automatic differentiation remains near 21-22 dB across the entire range tested (Fig. 5c). With one to six raw images, one to two orders of magnitude fewer than the 600 used for the ground-truth reference, Gaussian splatting already recovers cellular morphology close to the densely sampled result. Automatic differentiation contains structured artefacts that are suppressed only with dense sampling (Fig. 5d,e). The experimental reference is itself a reconstruction. The ground-truth simulation in Supplementary Fig. S12 therefore provides the unbiased form of this comparison, and it shows the same ordering.

The same behaviour holds for a structurally very different specimen. Supplementary Fig. S7 shows a gigapixel coded reconstruction of $Na_2CO_3$ crystals at 10.68 bits per pixel: the primitive distribution sharpens along the crystalline striations as optimization proceeds, Gaussian splatting resolves the striations from 6 raw measurements, and automatic differentiation requires substantially more data before the phase structure emerges. On a label-free thyroid smear, Gaussian splatting produces structurally consistent amplitude and phase estimates from as few as 5 measurements, a budget at which automatic differentiation is dominated by noise (Supplementary Fig. S13).

**Pupil-plane probe recovery for conventional optical, X-ray and electron ptychography**

In Fourier ptychography the unknown part of the imaging system is the pupil, and Figs. 3 and 4 showed that a primitive population in the pupil plane recovers it from a random start where pixel-grid solvers fail. Conventional ptychography poses a complementary problem. The unknown is the probe, a localized beam scanned across the specimen, and solvers customarily represent it on the sample-plane pixel grid and initialize it from a physical model of the beam. Yet the probe is itself the product of optics. A lens, a zone plate or a condenser aperture forms it, so its natural description is its pupil, the aperture-plane field from which the real-space beam follows by a Fourier transform. We therefore represent the probe of conventional ptychography by a second primitive population in its pupil plane, exactly as the pupil in Fourier ptychography, and recover object and probe jointly from a random pupil. No aperture size, defocus, zone-plate geometry or convergence angle is supplied (Methods).

We tested this at three wavelengths. Figure 6a,b shows conventional optical ptychography of a plant leaf section at $\lambda$ = 675 nm under a structured illumination beam, with random jitter added to the scan positions; Fig. 6c,d shows X-ray ptychography of gold nanoparticles at $\lambda$ = 1.24 nm[57]; and Fig. 6e,f shows ultralow-dose electron ptychography of a methylammonium lead iodide ($MAPbI_3$) nanosheet acquired by event-driven 4D-STEM at 200 keV[58]. The optical reconstruction resolves the cellular architecture of the section, and the probe converges to a structured beam confined to a circular pupil (Fig. 6b). The X-ray reconstruction resolves individual nanoparticles and their clusters, and the probe converges to the zone-plate focus, with the annular pupil of the zone plate appearing in reciprocal space (Fig. 6d). The electron reconstruction resolves the perovskite lattice at a dose of tens of electrons per square ångström, and the probe converges to an aperture-limited focused probe whose pupil is the disk of the probe-forming aperture (Fig. 6f). In each case the pupil that emerges is the one the optics would predict, although nothing about the optics was supplied. Supplementary Fig. S14 compares Gaussian splatting with the reference reconstruction of each dataset. The reference is initialized from the measured beam, including its bright-field disk, whereas Gaussian splatting starts from a random pupil.

The extension follows from representing the probe in reciprocal rather than real space. In the sample plane (real space), a speckled optical beam, a zone-plate focus and a focused electron probe have nothing in common; in the pupil plane (reciprocal space), each is a bounded field that a primitive population recovers from a random start by the mechanism of Figs. 3 and 4. One representation and one rasterizer therefore serve five orders of magnitude in wavelength with no model of the probe, and the electron case does so in a dose regime relevant to beam-sensitive materials.

## Discussion

We have introduced Gaussian-splatting ptychography, a reconstruction framework in which the specimen and probe are represented as explicit anisotropic Gaussian primitives rather than as values on a pixel grid. The framework is compact, storing gigapixel complex fields at 4.67 to 10.68 bits per pixel, a 6- to 14-fold reduction relative to an uncompressed dense complex-valued array. It is data-efficient, recovering cellular and crystalline morphology from a handful of raw measurements. It is also interpretable: every parameter is a position, a shape or a complex weight, and the primitive distribution itself traces the specimen morphology without segmentation.

The more consequential finding is that the reparameterization changes what can be recovered at all. Uniform phase transfer and robust blind pupil recovery are two manifestations of one structural property. In both cases the measurements leave certain modes weakly constrained. On a pixel grid those modes drift freely, because every pixel is an independent unknown. For the object, this appears as low-spatial-frequency phase collapsing toward the background. For the probe or pupil, it appears as an estimate that fills with pixel-scale noise until it fails to converge. In the splatting representation, neither outcome is readily expressible. Every pixel is written by a superposition of overlapping primitive footprints, and a weakly constrained mode is spanned by far fewer degrees of freedom.

The reparameterization leaves the loss as a function of the rendered field untouched; what changes is which fields can be produced and how each gradient step moves. A primitive affects the rendered field only through its footprint: changing its weight adds the footprint itself, and changing its position, scale or rotation adds a spatial derivative of it.

Both are smooth and extended patches, so no parameter can alter a single pixel on its own. Each step is in effect an adaptive smoothing of the pixel-space gradient with a kernel set by the learned covariances. Pixel-scale error modes cannot be expressed by such patches and average down over each footprint, whereas structure aligned with the primitives passes unattenuated. The failure modes of grid-based recovery are therefore not avoided by tuning; they are difficult to represent.

This framing also clarifies the relation to established remedies. An explicit smoothness or sparsity penalty on a pixel grid[29, 30, 31, 59] suppresses some of the weakly constrained modes, and a parametric aberration basis such as a Zernike expansion constrains the pupil to the smooth, low-order functions expected of a well-corrected objective. The differences are nonetheless substantive. A penalty adds a weight that must be balanced against a specimen-dependent loss scale, whereas here the loss is untouched. A parametric basis encodes a physical prior on the form of the aberration and cannot represent a field that departs from it. In contrast, the same primitive population that keeps an optical pupil smooth also represents the highly structured probes of conventional ptychography (insets of Fig. 6b2,d2), with no change to the representation or the rasterizer.

This restriction has a cost. A primitive that drifts away from useful support receives a vanishing gradient and is lost to the reconstruction, the origin of the localized convergence failure in Supplementary Fig. S3. Relocation repairs exactly the failure mode that the reparameterization introduces, and in doing so also improves the attainable resolution (Supplementary Fig. S4).

The framework is complementary to implicit neural representations, which likewise decouple the parameter count from the pixel grid[18, 36, 37, 38, 39, 40]. The distinction is that a neural field encodes the object in network weights that carry no individual physical meaning. Each Gaussian primitive, by contrast, is a localized, oriented element of the field that can be inspected, counted and, in principle, initialized from prior knowledge. The same property makes the reconstruction directly usable after optimization. The primitives form a point cloud with attributes, and operations that would require a separate image-analysis pass on a pixel grid become operations on that cloud: selecting primitives by position, density or weight segments regions; grouping them by orientation yields a fibre-alignment field; removing a class of primitives, such as a slowly varying background, edits the field without re-running the reconstruction. The primitive maps in Fig. 3 already trace glandular boundaries and stromal orientation in this way, before any pixel of the rendered image has been analysed.

Because primitives are countable, the compression curve acquires a meaning that a parameter count does not. The knee in Supplementary Figs. S5 and S6 can be read as an estimate of the specimen's effective information content, the number of primitives needed to describe the object rather than the number of pixels needed to sample it. On this reading, an anisotropic Gaussian is a minimum-uncertainty element jointly localized in space and spatial frequency[60], and optimization amounts to an adaptive tiling of the region of phase space that the specimen actually occupies, in contrast to the uniform tiling imposed by a pixel or plane-wave basis.

Three limitations point to immediate extensions. First, Gaussians are not band-limited, so their spectral tails extend beyond any physical pupil; replacing the Gaussian kernel with the point-spread function of the instrument, such as the Airy pattern of a circular aperture, would match the atom to the imaging system. Second, the reconstructions use a single-slice multiplicative model; optically thick specimens beyond the projection approximation would require multislice propagation[61, 62]. Third, the primitive count and scale range are chosen per modality, much as the update step sizes and initial probe models of pixel-grid solvers are, but unlike those they refer to the specimen rather than to the optimizer. The compactness sweeps of Supplementary Figs. S5 and S6 show that the useful count tracks the structural complexity of the specimen, so a rule that derives both settings from a coarse pilot reconstruction and the field size is a natural future direction.

Other extensions follow from the representation itself. Three-dimensional Gaussian primitives, the setting for which splatting was devised, would carry the parameterization to Fourier ptychographic diffraction tomography[63, 64] and to time-varying specimens[18]. More broadly, the argument is not specific to ptychography. Wherever an inverse problem is solved on a dense grid with weakly constrained modes and expensive measurements, an explicit primitive representation may change not only the cost of the reconstruction but also its content. Early applications to X-ray tomography[65], cryo-electron microscopy[66] and in-line holography[46] point the same way.

# Methods

**Gaussian primitives with real-imaginary complex weights.** The complex field is represented by $N$ anisotropic two-dimensional Gaussian primitives. Each primitive carries seven learnable scalars, namely a centre, two scales and an in-plane rotation that define its covariance, and a complex weight stored as real and imaginary components. The rendered field is the coherent sum of all primitive footprints, with the real and imaginary channels accumulated separately. We avoid an amplitude-phase weight because the phase is periodic: its branch cut at $\pm\pi$ cannot be crossed by a gradient step, and its gradient scales with the amplitude and vanishes for exactly the weak primitives that relocation acts on. The real-imaginary form spans the identical set of fields under the identical loss, is linear in both components, and has no such discontinuities (Supplementary Note 1, Supplementary Fig. S1).

Primitive scales are confined to a fixed interval. The lower bound imposes a band limit directly in the representation and prevents collapse toward delta-function support, while the upper bound prevents any primitive from acting as a global background term. Rendering is performed by a custom tile-based CUDA rasterizer with analytic gradients, which enters the optimization as a single differentiable operator.

The probe or pupil is carried by a second primitive population rendered by the same rasterizer, with its own count and scale range. In Fourier ptychography, the pupil phase is represented by primitives, so the recovered quantity is the pupil aberration. In the conventional optical, X-ray and electron experiments of Fig. 6, the probe is represented in the same way: the primitives render a complex pupil function whose inverse Fourier transform is the real-space probe, normalized to unit energy at every forward pass. The pupil primitives start from the same state as the object primitives: positions drawn uniformly at random over the canvas and identical complex weights, so the initial pupil phase is flat and carries no aberration information. No aperture size, defocus, zone-plate geometry or convergence angle is supplied.

**Forward models.** For lensless coded imaging, the rendered complex object is propagated by the angular-spectrum method from the object to the coded-surface plane, and multiplied by the laterally shifted complex coded-surface profile. The field is then propagated over the coded-surface-to-detector distance and downsampled by the pixel-super-resolution factor to match the detector sampling. The coded surface is pre-calibrated on a weakly scattering reference specimen and held fixed during reconstruction. For lens-based Fourier ptychography, the rendered object spectrum is cropped by the pupil aperture at the illumination-dependent offset determined by the LED position, multiplied by the current pupil estimate, and inverse-transformed to give the predicted low-resolution intensity. Illumination wavevectors are computed from the measured LED array geometry, including the refractive offset of the substrate. The resulting sub-aperture centres are verified against the numerical-aperture cutoff before reconstruction. For conventional ptychography, the rendered object is multiplied by the shifted probe and propagated to the far field. The three forward models are written out in Supplementary Note 4 (equations S6 to S8).

**Optimization and primitive relocation.** All parameters are optimized by Adam[67] with per-parameter groups, so that the optimizer state can be reset selectively after relocation, and with a cosine-annealed learning-rate schedule. Fourier ptychographic and coded-ptychography reconstructions use a gradient-domain loss[68], the per-pixel magnitude of the gradient of the amplitude residual averaged over the image; conventional ptychography uses a smooth-L1 loss on intensities (Supplementary Note 4, equations S9 to S11). Relocation follows the sampling-based reformulation of splatting optimization[43]. The primitive count is fixed throughout. At regular intervals the primitives are ranked by the magnitude of their complex weight, and a preset fraction of the lowest-ranked primitives, together with any primitive whose scale has saturated at a bound, is declared dead. Each dead primitive is respawned beside a surviving primitive sampled in proportion to its weight, and the weight is shared so that the expected rendered field is unchanged at the moment of relocation; the optimizer state of the relocated primitive is reset. The hyperparameters of the released demonstration configurations are listed in Supplementary Table 2, and the implementation is described in Supplementary Notes 1-5. Ablations with relocation disabled are reported in Supplementary Figs. S3 and S4. Other experimental parameters like scan positions can be included in the same optimization scheme. The scan-position result of Supplementary Fig. S2 uses a simulated dataset with known ground-truth positions.

**Lensless coded ptychography.** We acquired the coded-ptychography data (Fig. 5 and Supplementary Figs. S1, S3, S7 and S13) with a coded-sensor platform. The coded surface, a disorder-engineered layer of sub-micron intensity and phase scatterers, is fabricated directly on the protective glass of a CMOS image sensor (Sony IMX226, 1.85 µm pixel pitch), facing the pixels, so that it never contacts the specimen and is robust to contamination and mechanical wear. A 405-nm laser diode (Thorlabs LP405-SF10, about 10 mW) provides coherent illumination across the full sensor area, and the specimen is placed 0.2-2 mm above the coded surface. The complex transmission profile of the coded surface is calibrated once and is held fixed in all subsequent reconstructions. For each dataset, the coded sensor is translated laterally beneath the specimen through a grid of positions. Live cultures were imaged with the entire platform placed inside the incubator.

**Fourier ptychography.** We acquired the Fourier ptychographic data (Figs. 2-4 and Supplementary Figs. S10 and S11) using a regular upright microscope fitted with a 2×, 0.1-NA objective and a programmable LED array with 2.5 mm pitch, placed 50-70 mm below the specimen. The outermost LEDs used correspond to an illumination numerical aperture of about 0.4, giving a synthetic numerical aperture of about 0.5. The illumination wavelengths were 630, 530 and 475 nm (red, green and blue). LEDs were switched on one at a time and one image was recorded per LED. The full acquisitions used a 13 × 13 LED grid. The reduced-overlap experiments of Fig. 4 used 169, 49 and 25 LEDs on 13 × 13, 7 × 7 and 5 × 5 grids centred on the optical axis, corresponding to approximately 70%, 41% and 17% aperture overlap. Illumination wavevectors were computed from the LED positions and the refractive offset of the glass substrate. A single learnable quadratic phase term is fitted inside the Fourier ptychographic forward model to absorb the residual curvature of the LED illumination (Supplementary Note 4)[69]; the same term is used in the pixel-grid reference reconstructions, so that the comparison of low-frequency phase is unaffected by it.

**Conventional optical, X-ray and electron ptychography.** We built a scanning conventional-ptychography setup for this study and used it to image the plant section of Fig. 6a,b at a wavelength of 675 nm. The illumination was a structured probe, and the diffraction patterns were recorded on a camera downstream of the specimen, with an effective specimen-to-detector distance of 29.5 mm and an effective detector pixel size of 29.6 µm. The specimen was scanned on a 10 × 20 grid with nominal step sizes of approximately 80 µm along the x axis and 40 µm along the y axis, with random position jitter of up to ~16 µm added to each scan point, giving 200 diffraction patterns of 512 × 512 pixels. The structured probe makes this dataset a demanding test of joint object-probe recovery: no measurement or model of the probe was supplied to the reconstruction, which recovered it blind as Gaussian primitives in the detector's reciprocal space.

The X-ray gold-nanoparticle dataset of Fig. 6c,d and Supplementary Fig. S4 ($\lambda = 1.24$ nm, 1 keV) is the reference dataset distributed with the SHARP solver[57]. For the Gaussian-splatting reconstructions, only the diffraction patterns and nominal scan positions were used. The electron ptychography data were acquired on an aberration-corrected Thermo Fisher Themis Z at 200 keV with an event-driven Timepix3 direct-detection camera at 500-ns dwell time on a 512 × 512 probe-position grid, at a dose of approximately 11 $e^{-}$ $Å^{-2}$ per scan[58]. At this dose each diffraction pattern contains only a few electrons.

**Specimen preparation.** Bone-marrow mesenchymal stem cells (RASMX-01001, Cyagen) were cultured in growth medium (RASMX-90011, Cyagen) supplemented with 10% fetal bovine serum, seeded in standard culture dishes and imaged in the dish. The bacterial microcolony of Supplementary Fig. S1 was *Escherichia coli* ATCC 25922 grown in Mueller-Hinton broth and imaged on an agar pad. $Na_2CO_3$ crystals were grown by drying a droplet of saturated aqueous sodium carbonate solution on a glass slide at room temperature; the same preparation was imaged by Fourier ptychography (Fig. 2d) and by coded ptychography (Supplementary Fig. S7). Cystine crystals were grown by slow evaporation of a saturated aqueous L-cystine solution on a glass slide. The quantitative phase target of Fig. 2a was a commercial target with etched features of nominal height 300 nm (Benchmark Technologies); this height is the ground truth against which the recovered step in Fig. 2a is compared. The blood smear used as the calibration reference and as the specimen of Fig. 4 and Supplementary Fig. S11 was a commercially prepared, Wright-stained human blood smear slide (Carolina Biological Supply). The thyroid smear of Supplementary Figs. S1 and S13 was a de-identified, unstained fine-needle-aspiration smear (Pathware Inc); no patient-identifiable information was accessible to the authors, and the use of these de-identified specimens was determined not to constitute human-subjects research. The mouse-kidney section of Fig. 2c and the H&E-stained section of Fig. 3 were commercially prepared paraffin sections (Carolina Biological Supply).

**Image quality assessment.** Reconstruction quality is quantified by PSNR against the ground truth in simulation, and against a densely sampled reconstruction from 600 raw measurements in the coded-ptychography experiments. Amplitude and phase are evaluated separately. Phase images are compared after removal of a global piston term. Resolution in X-ray ptychography (Supplementary Fig. S4) is estimated by Fourier ring correlation between two independent reconstructions using the 1/2-bit threshold[70].

## Data availability

The experimental datasets that accompany the code implementations are available in Supplementary Note 5 with a Zenodo repository link.

## Code availability

The code implementation of Gaussian-splatting ptychography, including demonstration notebooks for lensless and lens-based ptychographic implementations, is available in Supplementary Note 5 with a Zenodo repository link.

## Acknowledgements

This work was partially supported by the Department of Energy SC0025582 and the National Institutes of Health R01-EB034744. The content of the article does not necessarily reflect the position or policy of the US government, and no official endorsement should be inferred. Q. Z. acknowledges the support of the G. E. fellowship.

## Author contributions

Q. Z. and G. Z. conceived the original concept and supervised the project. Q. Z. and Z. H. developed the Gaussian-splatting ptychography framework and implemented the reconstruction code. Q. Z., A. M., Z. Z. and R. W. built the lensless coded, Fourier and conventional ptychographic platforms and acquired all optical data reported here. Q. Z., Z. H. and Z. Z. performed the reconstructions and prepared the display items. R. W. prepared the specimens. Q. Z. prepared the Supplementary Information. D. B., C. Y., A. M., D. G., M. L., C. A., and A. B. participated in the discussion and interpretation of the results. Q. Z., Z. H. and G. Z. wrote the manuscript with input from all authors.

## Competing interests

G. Z. is a named inventor of related patents on Fourier ptychography and coded ptychography. C. Y. is a named inventor of related patents on Fourier ptychography. A. M. is a named inventor of related patents on conventional ptychography. The other authors declare no competing interests.

## Supplementary information

Supplementary Fig. S1 | Real-imaginary weights avoid the $2\pi$ wrapping discontinuities of amplitude-phase weights and stabilize Gaussian-splatting reconstruction across diverse specimens.
Supplementary Fig. S2 | Joint scan-position correction in Gaussian-splatting ptychography.
Supplementary Fig. S3 | Primitive relocation aligns Gaussians with sample structure and avoids local convergence failure in coded ptychography.
Supplementary Fig. S4 | Primitive relocation improves half-period resolution from 22.7 to 14.1 nm in conventional X-ray ptychography.
Supplementary Fig. S5 | Reconstruction quality versus representation compactness in Fourier ptychography.

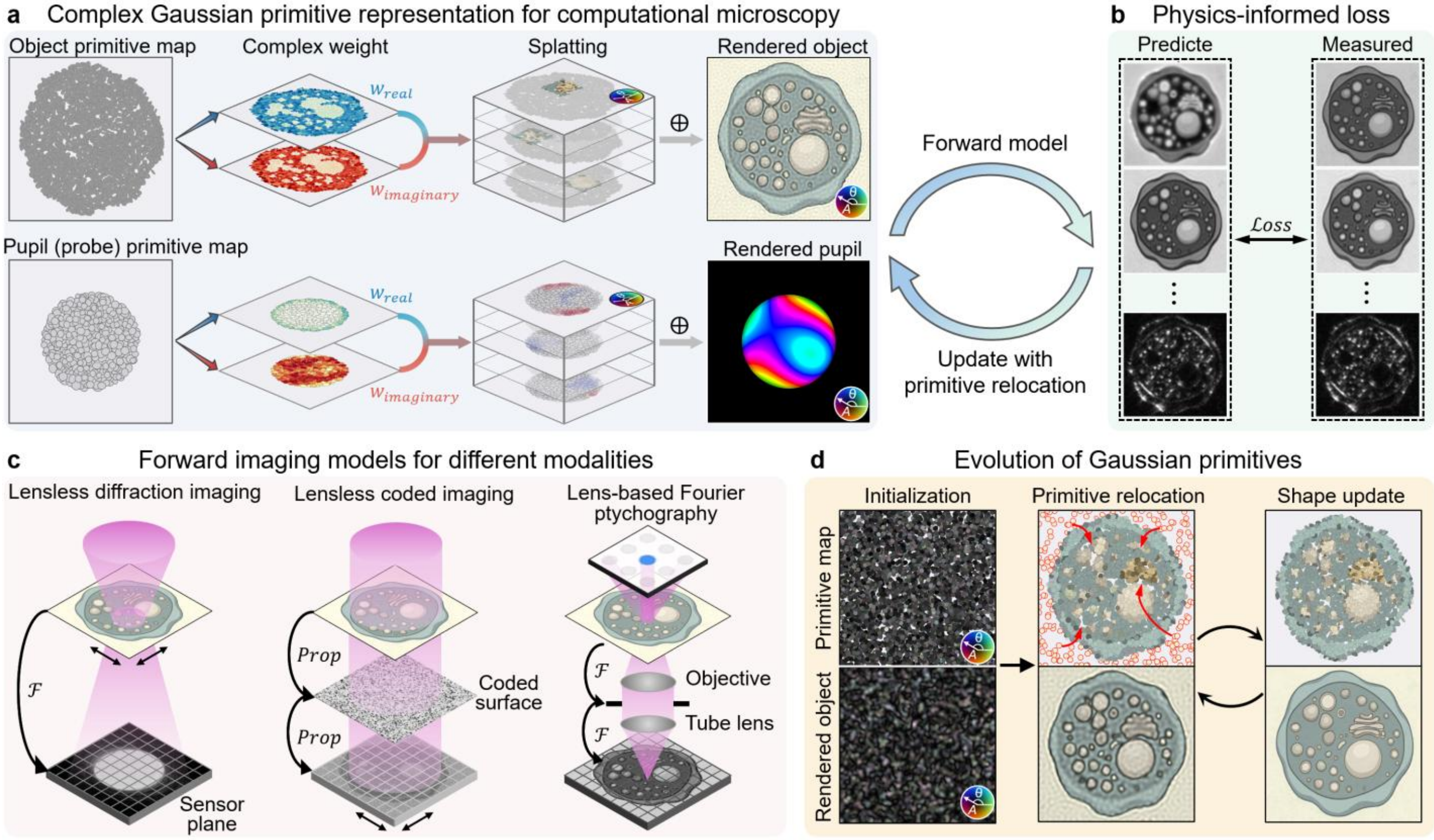

**Fig. 1 | Gaussian-splatting representation for computational microscopy. a,** Complex Gaussian primitive representation of the object and probe or pupil fields. Each field is parameterized by anisotropic Gaussian primitives with learnable positions, shapes, orientations and complex weights. Each complex weight is stored as real and imaginary components, which are splatted onto a regular grid and summed to render the complex object and pupil. **b,** Physics-informed loss. The rendered fields are propagated through a modality-specific forward model to generate predicted measurements, and the loss is computed between the predicted and measured intensity stacks. The primitive weights, shapes and locations are updated jointly during optimization. **c,** Forward imaging models for the three modalities. Conventional ptychography records the far-field diffraction of a scanned probe at the sensor plane. Lensless coded imaging introduces a coded surface in the propagation path to modulate the transmitted wavefield. Lens-based Fourier ptychography acquires band-limited intensity images under angle-varied illumination through an objective-lens system. **d,** Evolution of the Gaussian primitives. Starting from a random initialization, primitive relocation reallocates primitives toward informative sample regions, and subsequent shape updates produce a compact anisotropic representation aligned with the sample morphology, yielding a structured complex-field reconstruction.

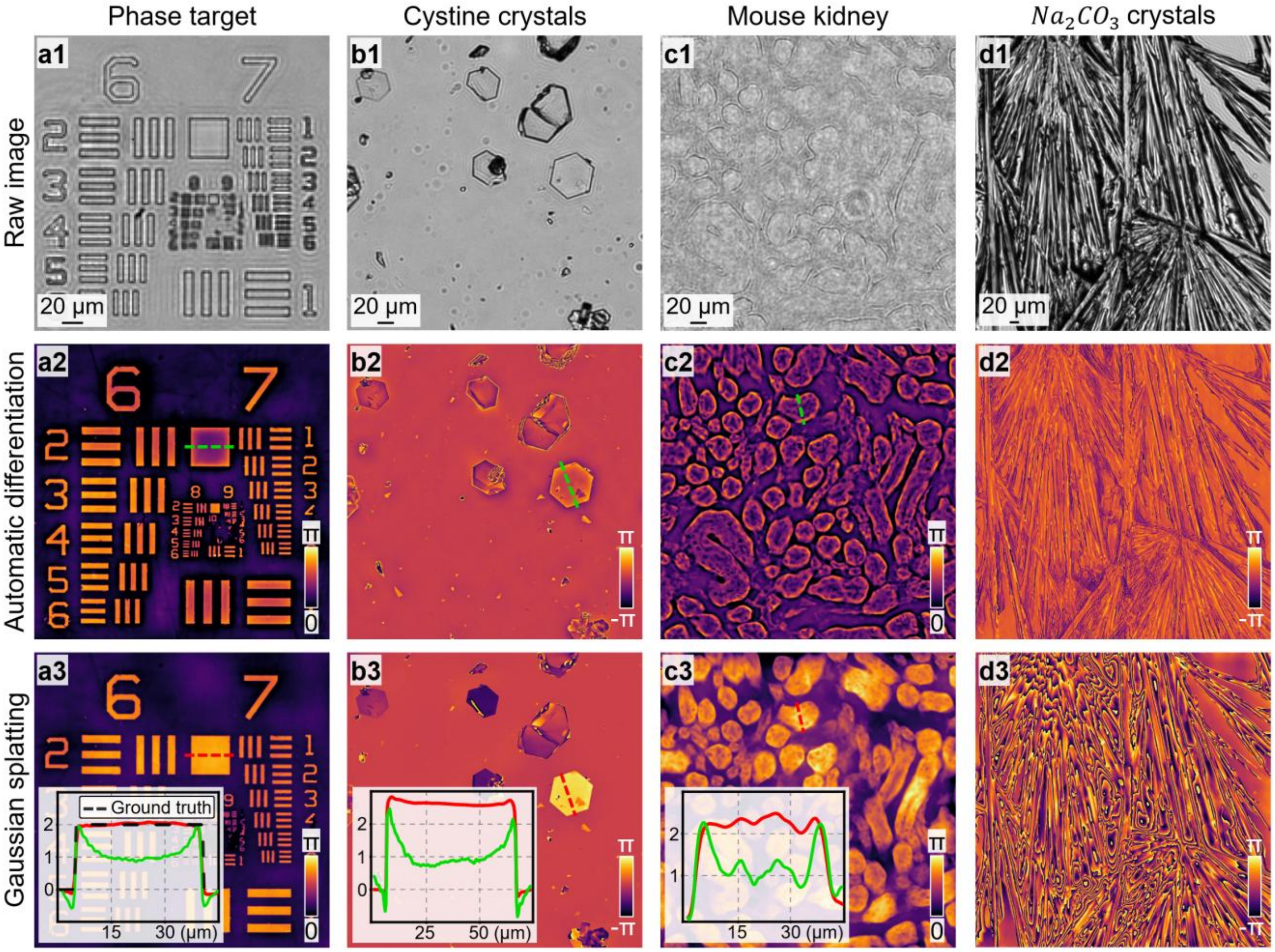

**Fig. 2 | Uniform phase transfer across spatial frequencies with Gaussian-splatting rendering. a1–d1,** Representative raw measurements of four specimens acquired by Fourier ptychographic microscopy. **a2–d2,** Phase reconstructions obtained by pixel-grid automatic differentiation. **a3–d3,** Corresponding phase reconstructions obtained by Gaussian splatting from the same data, forward model, loss and optimizer. Insets in **a3–c3** plot the phase profiles along the dashed lines marked in **a2–c2** (green, automatic differentiation) and **a3–c3** (red, Gaussian splatting), with the ground-truth profile of the phase target overlaid in **a3** (black dashed). Automatic differentiation resolves edges but loses low-spatial-frequency phase, so the interiors of extended features sag toward the background level and the recovered step height falls well below the true value. Gaussian splatting reproduces the full step height, quantitatively matching the manufactured value, and maintains a flat response across each feature, demonstrating uniform phase transfer from high spatial frequencies down to the low-frequency limit. In **d2–d3**, the same low-frequency loss leaves the automatic-differentiation reconstruction with little more than the striation edges, whereas Gaussian splatting recovers the accumulated optical thickness, which appears as dense 2π phase-wrapping fringes.

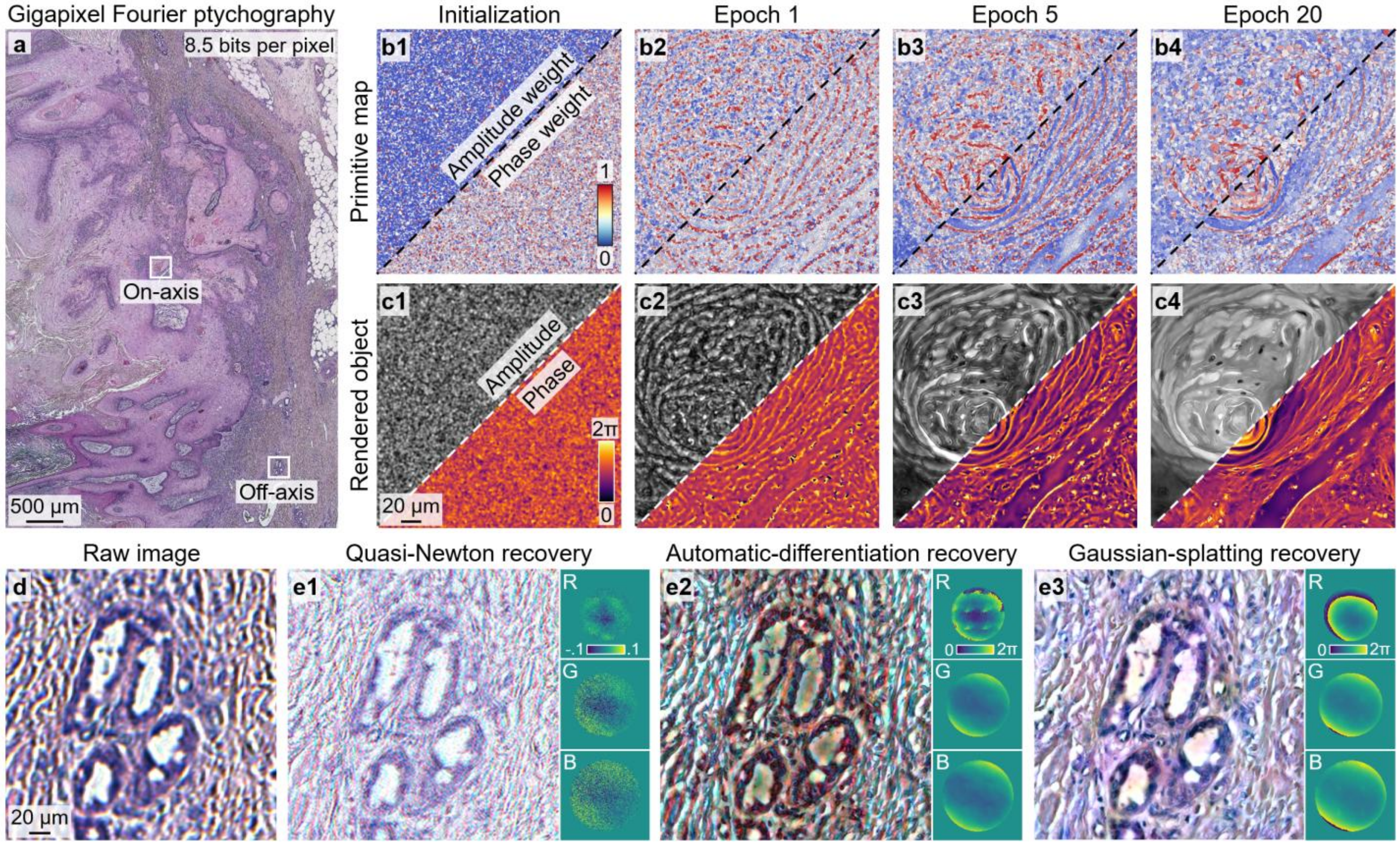

**Fig. 3 | Gaussian splatting enables blind pupil recovery in gigapixel Fourier ptychography. a,** Full-field colour reconstruction of an H&E-stained tissue section, recovered by the Gaussian-primitive representation at 8.5 bits per pixel per colour channel. Boxes mark the on-axis and off-axis regions of interest. **b,** Evolution of the Gaussian primitive map for the green channel of the on-axis region at initialization and after 1, 5 and 20 epochs, with the amplitude weight in the upper-left triangle and the phase weight in the lower-right triangle. **c,** Corresponding rendered complex object at the same epochs, with amplitude in the upper-left triangle and phase in the lower-right triangle. The primitives condense from a random distribution onto the tissue architecture, and the rendered object evolves from noise to a high-contrast complex-field reconstruction. **d,** Low-resolution raw measurement from the off-axis region. **e,** Reconstructions of the same region by quasi-Newton optimization, pixel-grid automatic differentiation and Gaussian splatting, each with the recovered pupil phase for the red, green and blue channels shown as insets. All three methods recover the pupil as a free-form function from the same raw data, with no parametric aberration basis and no prior on the form of the aberration. Gaussian splatting recovers consistent, smoothly varying aberrations across all three channels and resolves glandular morphology with the highest fidelity, demonstrating that the primitive representation stabilizes blind aberration recovery where pixel-grid solvers do not.

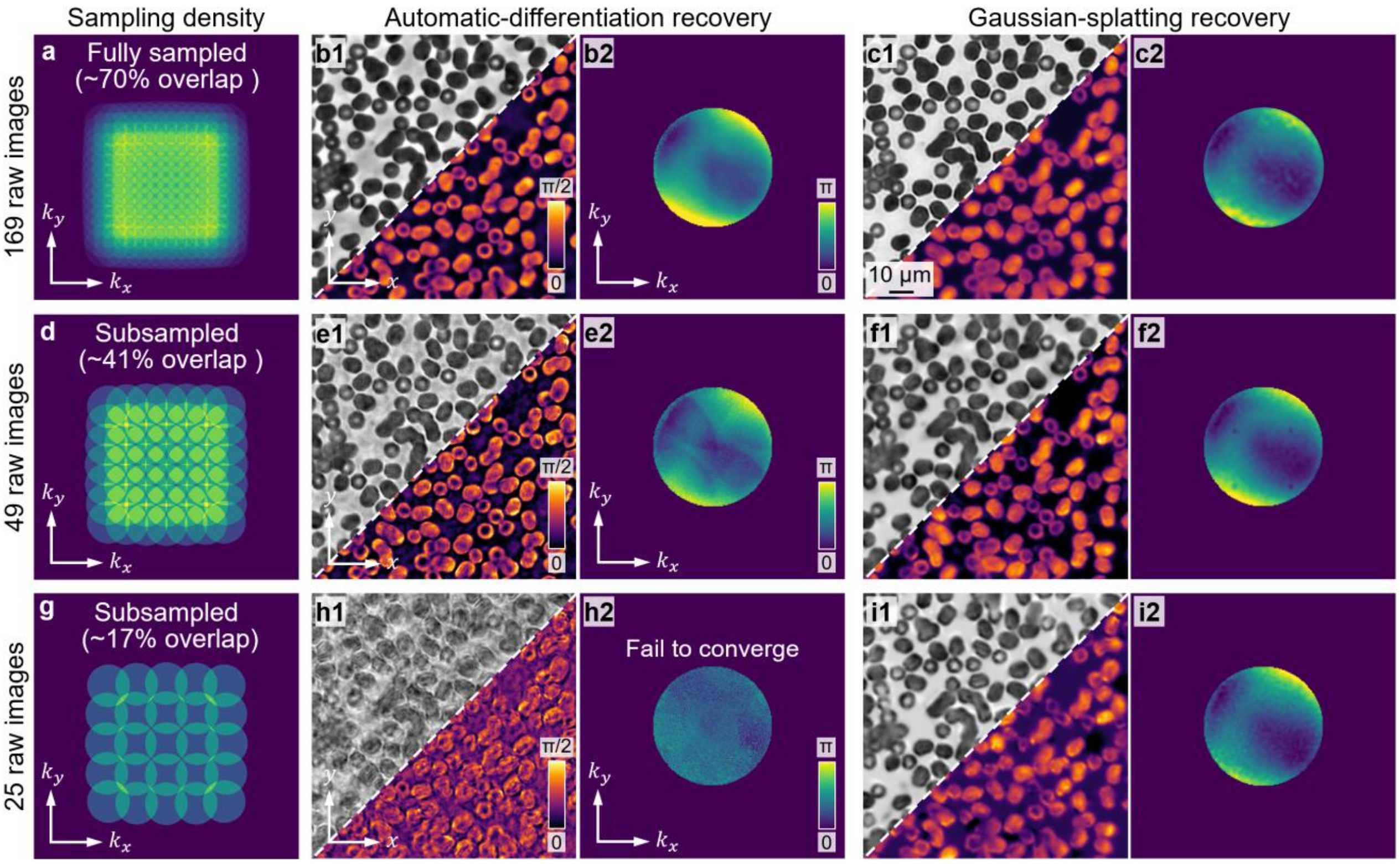

**Fig. 4 | Fourier ptychographic reconstruction of blood cells under reduced aperture overlap. a,d,g,** Fourier-space sampling densities for 169, 49 and 25 raw measurements, corresponding to approximately 70%, 41% and 17% aperture overlap, respectively. **b,e,h,** Object and pupil reconstructions obtained by pixel-grid automatic differentiation. **c,f,i,** Corresponding reconstructions obtained by Gaussian splatting. For each reconstruction pair, the first panel shows the recovered object with amplitude in the upper-left triangle and phase in the lower-right triangle, and the second panel shows the recovered pupil aberration. At 169 measurements, both methods recover cellular morphology and a smooth pupil aberration. As the aperture overlap decreases, the automatic-differentiation reconstruction degrades progressively and fails to converge at 25 measurements, whereas Gaussian splatting preserves cellular morphology and recovers a pupil aberration consistent with the higher-overlap cases.

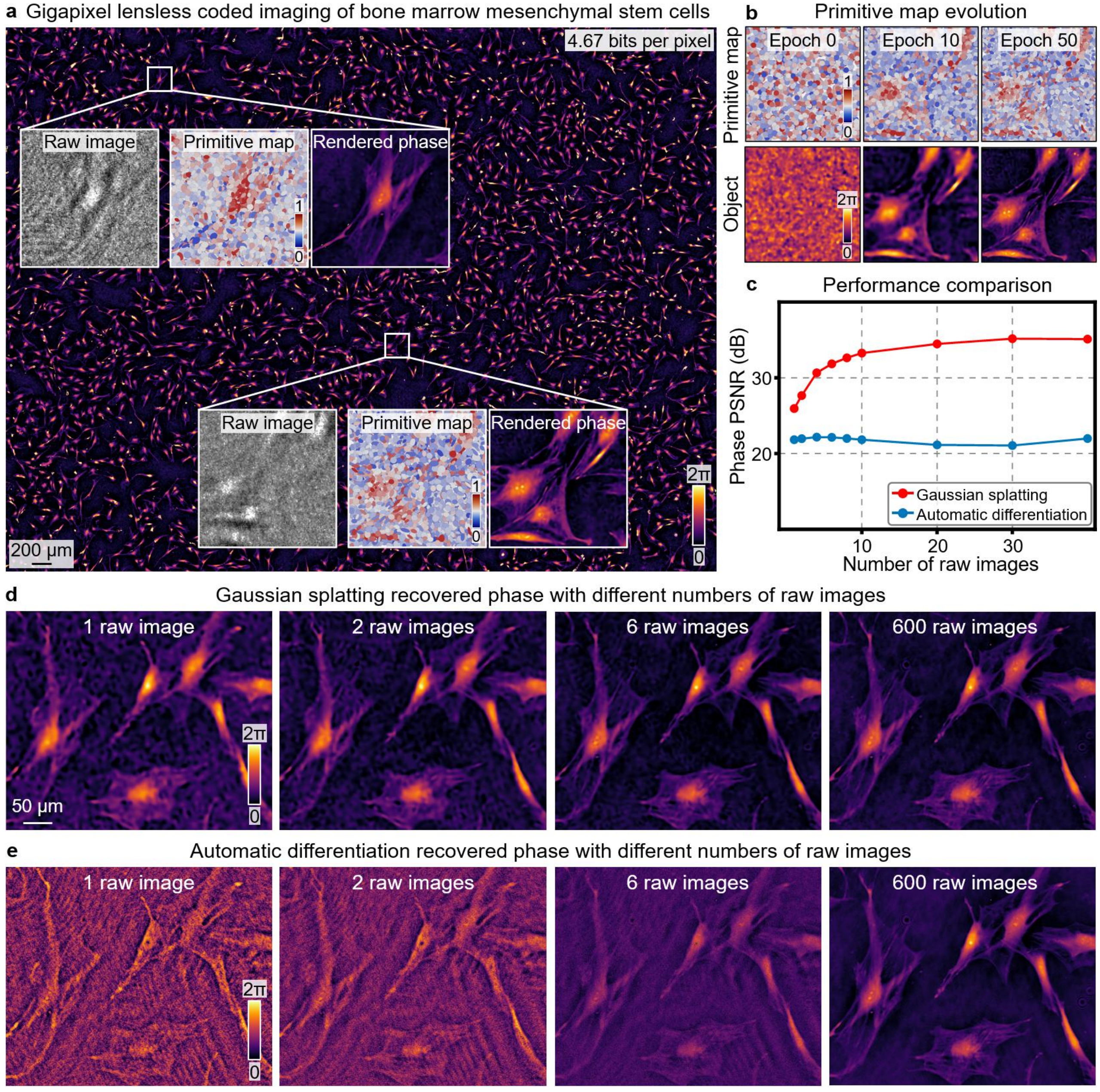

**Fig. 5 | Gigapixel lensless coded imaging of bone-marrow mesenchymal stem cells. a,** Full-field phase reconstruction of bone-marrow mesenchymal stem cells, stored by the Gaussian-primitive representation at 4.67 bits per pixel. Insets show representative raw measurements, primitive phase maps and rendered phase reconstructions from two regions of interest. **b,** Evolution of the primitive phase maps and the corresponding rendered phases at epochs 0, 10 and 50. The primitives progressively organize along cell bodies and protrusions, yielding sharper phase features as optimization proceeds. **c,** Phase PSNR of Gaussian splatting and pixel-grid automatic differentiation as a function of the number of raw measurements, using the 600-measurement reconstruction as the reference. Gaussian splatting maintains a higher phase PSNR under sparse measurements. **d,** Gaussian-splatting phase reconstructions recovered from 1, 2, 6 and 600 raw measurements. **e,** Automatic-differentiation phase reconstructions under the same measurement budgets. With one to six measurements, Gaussian splatting recovers cellular morphology close to the densely sampled result, whereas automatic differentiation contains structured artefacts that are reduced only with dense sampling.

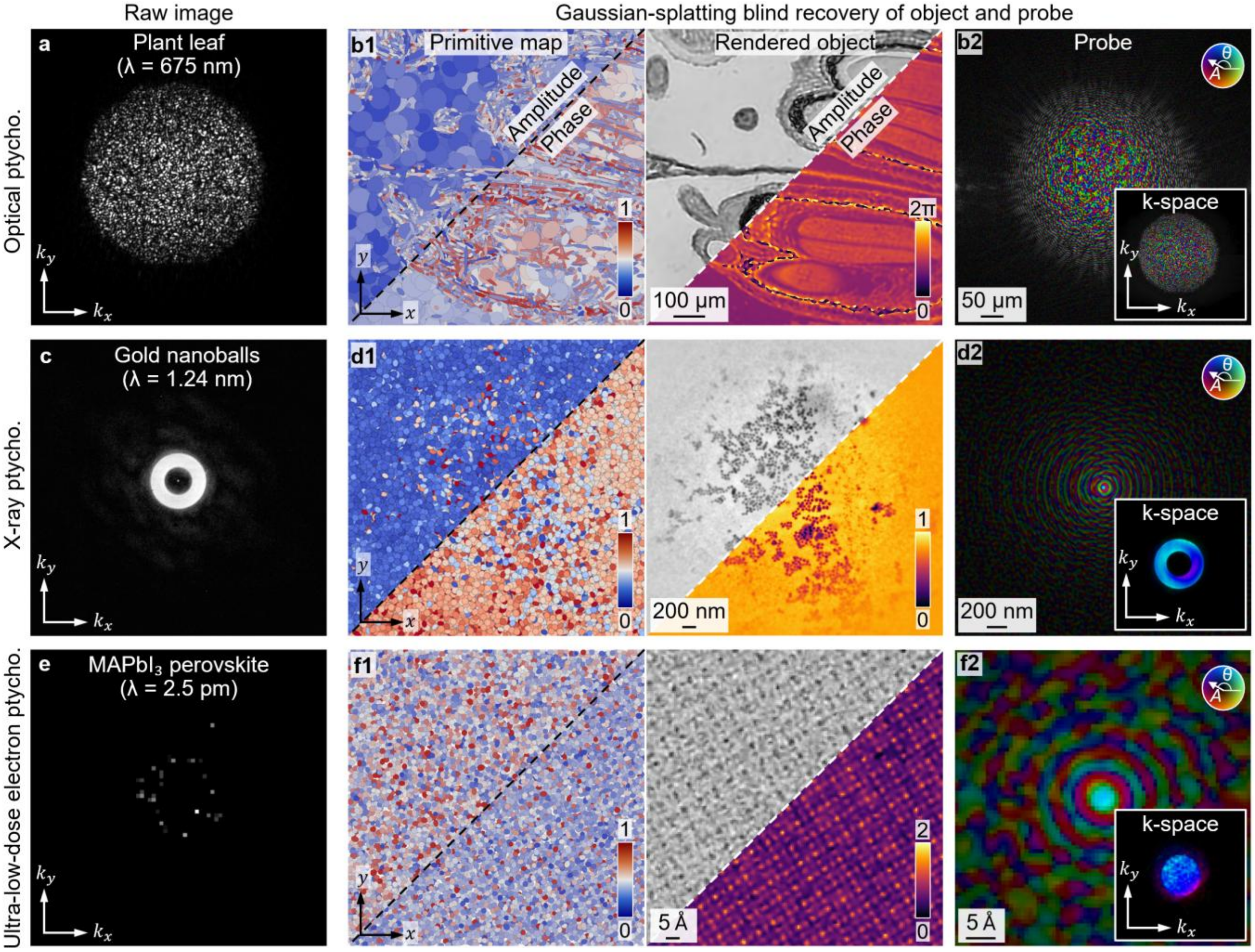

**Fig. 6 | Pupil-plane probe recovery in optical, X-ray and electron ptychography.** In all panels the probe is represented by Gaussian primitives in its pupil plane and recovered jointly with the object from a random pupil; no aperture size, defocus, zone-plate geometry or convergence angle is supplied. **a, b**, Conventional optical ptychography of a plant leaf section at λ = 675 nm under a structured illumination beam. **a**, One raw diffraction pattern. **b1**, Primitive map (magnitude and phase of the complex weight, each normalized to 0 to 1) and rendered object (amplitude, upper left; phase, lower right). **b2**, Recovered real-space probe, shown as amplitude (brightness) and phase (hue). Inset, the recovered pupil-plane field (k-space) from which the probe is rendered: a structured beam confined to a circular pupil. **c, d**, X-ray ptychography of gold nanoparticles at λ = 1.24 nm (1 keV)[57]; panels as in **a, b**. The recovered probe (**d2**) is the zone-plate focus and its pupil (inset) the annulus of the zone plate. **e, f**, Ultralow-dose electron ptychography of a $MAPbI_3$ nanosheet acquired by event-driven 4D-STEM at 200 keV (λ = 2.5 pm)[58]; panels as in **a, b**. At this dose the raw pattern in **e** contains only a few detected electrons, yet the rendered phase resolves the atomic lattice, the recovered probe (**f2**) shows the ring structure of an aperture-limited focused probe, and its pupil (inset) is the disk of the probe-forming aperture. Comparisons with the default reference reconstructions of the same data are shown in Supplementary Fig. S14.